\documentclass[%
 reprint,
 amsmath,amssymb,
 aps,
]{revtex4-1}

\usepackage{dcolumn}
\usepackage{bm}

\begin{document}

\def\be{\begin{equation}}
\def\ee{\end{equation}}
\def\bear{\begin{eqnarray}}
\def\eear{\end{eqnarray}}
\def\nn{\nonumber}

\newcommand\bra[1]{{\left\langle{#1}\right\rvert}} 
\newcommand\ket[1]{{\left\lvert{#1}\right\rangle}} 

\newcommand{\R}{\mathbb{R}}

\def\curveC{{C}}

\def\dPath{{P}}
\def\newPhi{{\widetilde{\Phi}}}

\def\Bz{{B}}
\def\Hz{{H}}

\newcommand\rep[1]{{\mathbf{#1}}} 
\newcommand\brep[1]{{\overline{\mathbf{#1}}}} 


\def\zz{{\mathbf{z}}}
\def\zone{{\mathfrak{1}}}
\def\ztwo{{\mathfrak{2}}}
\def\zthree{{\mathfrak{3}}}
\def\bzz{{\overline{\zz}}}
\def\bzone{{\overline{\zone}}}
\def\bztwo{{\overline{\ztwo}}}
\def\bzthree{{\overline{\zthree}}}

\def\sa{{\mathbf{a}}} 
\def\sb{{\mathbf{b}}} 
\def\sc{{\mathbf{c}}} 
\def\sd{{\mathbf{d}}} 
\def\se{{\mathbf{e}}} 
\def\sf{{\mathbf{f}}} 
\def\sg{{\mathbf{g}}} 
\def\sh{{\mathbf{h}}} 

\def\Pz{{\chi}} 
\def\Sz{{\varphi}} 
\def\aPq{{\psi}} 
\def\aSq{{\phi}} 
\def\bPq{{\overline{\aPq}}} 
\def\bSq{{\overline{\aSq}}} 

\def\pSUSY{{\eta}}

\def\ri{{i}} 
\def\rj{{j}} 
\def\rk{{k}} 
\def\rl{{l}} 

\def\pEE{{\mathfrak{V}}} 

\def\DCov{{{\mathcal D}}}
\newcommand\DCovDn[2]{{\DCov^{\lbrack{#1}\rbrack}_{#2}}} 
\newcommand\DCovUp[2]{{\DCov^{\lbrack{#1}\rbrack\,{#2}}}} 

\def\SCP{{\Sigma}} 

\newcommand\intSCP[1]{{\boxed{{#1}}\,}} 
\newcommand\rintSCP[1]{{\overline{\boxed{{#1}}}\,}} 

\def\zC{{\zeta}} 

\def\LagO{{\hat{\mathcal{L}}}}

\def\sx{{\mathbf{x}}} 
\def\sy{{\mathbf{y}}} 
\def\sz{{\mathbf{z}}} 
\def\su{{\mathbf{u}}} 
\def\sv{{\mathbf{v}}} 
\def\rx{{x}} 
\def\ry{{y}} 

\def\JTnew{{{J}}} 
\def\JS{{\mathfrak{J}}} 
\def\JP{{\mathfrak{K}}} 

\def\JSR{{\widetilde{\mathcal{J}}}} 

\def\QT{{M}} 

\def\aSX{{\mathfrak{S}}} 
\def\bSX{{\overline{\aSX}}} 

\def\cSX{{{S}}} 

\def\mV{{\mathbf{V}}} 
\def\mF{{\mathbf{F}}} 
\def\mR{{\rho}} 
\def\mL{{\lambda}} 

\def\yV{{\xi}} 

\def\dualJS{{\widetilde{\JS}}}

\def\wA{{\widetilde{A}}}
\def\wF{{\widetilde{F}}}
\def\wL{{\widetilde{L}}}

\def\tB{{\widetilde{B}}} 
\def\tH{{\widetilde{H}}} 
\def\tK{{\widetilde{K}}} 

\def\defineas{{\equiv}}

\def\lL{{l}} 

\def\AcI{{I}} 

\def\vy{{\mathbf y}} 
\def\rh{{\mathbf h}} 
\def\rR{{r}} 

\newcommand\TransverseSpace[1]{{\mathbf{W}_{{#1}}}}  

\def\lP{\ell_p} 


\title{%
Supersymmetric interactions of a six-dimensional self-dual tensor and fixed-shape second quantized strings
}

\author{Ori J. Ganor}
\email{ganor@berkeley.edu}
\affiliation{%
Department of Physics\\
366 LeConte Hall MC 7300\\
University of California\\
Berkeley, CA 94720
}%

\date{\today}

\begin{abstract}
``Curvepole $(2,0)$-theory'' is a deformation of the $(2,0)$-theory with nonlocal interactions.
A ``curvepole'' is defined as a two-dimensional generalization of a dipole.
It is an object of fixed two-dimensional shape whose boundary is a charged curve that interacts with a two-form gauge field.
Curvepole theory was previously only defined indirectly via M-theory.
Here we propose a supersymmetric Lagrangian, constructed explicitly up to quartic terms, for an ``abelian'' curvepole theory, which is an interacting deformation of the free $(2,0)$ tensor mutliplet.
This theory contains fields whose quanta are curvepoles (i.e., fixed-shape strings).
Supersymmetry is preserved (at least up to quartic terms) if the shape of the curvepoles is (2d) planar.
This nonlocal 6d QFT may also serve as a UV completion for certain (local) 5d gauge theories.
\end{abstract}

\maketitle



\section{\label{sec:intro}Introduction}

Gaining a better understanding of the 6d $(2,0)$-theory, originally discovered in \cite{Witten:1995zh}, has become a major objective in contemporary theoretical high energy physics.
The $(2,0)$-theory's relation to the low-energy limit of M$5$-branes \cite{Strominger:1995ac}, as well as its power to elucidate the strong coupling dynamics of 4d and lower dimensional gauge theories, has been realized at the outset \cite{Witten:1995zh}, and further developed in \cite{Ganor:1996xg, Cheung:1998te, Cheung:1998wj}, with a remarkably unified picture emerging more recently in \cite{Gaiotto:2009we,Alday:2009aq,Chacaltana:2010ks,Dimofte:2011ju,Gadde:2013sca,Aganagic:2015cta} and other works. These developments generate a strong motivation for finding a fundamental description of the $(2,0)$-theory.
Several promising approaches have been proposed over the years (see for example
\cite{Aharony:1997th,Aharony:1997an,ArkaniHamed:2001ie,Douglas:2010iu,Lambert:2010iw,Lambert:2010wm,Ho:2011ni,Huang:2012tu,Ho:2014eoa,Lambert:2016xbs,Kucharski:2017jwv}), but the problem is still open.
Part of the difficulty stems from the problem that (at least on the Coulomb branch) the theory describes an interacting anti-self-dual two-form gauge field, and the charged objects that couple to two-forms are strings rather than point particles.

It is therefore interesting to explore the various possibilities that arise from interacting two-form field theories, and in this letter we take first steps to construct a supersymmetric field theory describing a deformation of a free $(2,0)$ tensor multiplet (containing a two-form gauge field) where some of the scalars and spinors take the form of extended string-like objects whose boundary is charged under the two-form field. 

The sort of theory we are looking for was put forward in \cite{Dasgupta:2000ry} (referred to by another moniker) as an M-theoretic construction.
In order to better explain what kind of 6d field theory we seek to construct, it is useful to recall the ``dipole-theories'' in 4d.
In \cite{Bergman:2000cw,Bergman:2001rw,Dasgupta:2001zu}, motivated by the construction of Yang-Mills theory on a noncommutative space \cite{Connes:1997cr,Douglas:1997fm,Seiberg:1999vs}, a theory of ``fundamental'' dipoles interacting with a gauge field was constructed and embedded in string theory. The term ``fundamental'' here means that the dipoles of dipole-theory are quanta of a fundamental field, rather than composites of other fields. For example, a scalar fundamental dipole field $\phi$ couples to a gauge field $A_\mu dx^\mu$ via the covariant derivative 
\be\label{eqn:defDdipole}
D_\mu^{\lbrack\lL\rbrack}\phi(x)\defineas
\partial_\mu\phi(x)
+i q [A_\mu(x+\tfrac{1}{2}\lL)-A_\mu(x-\tfrac{1}{2}\lL)]\phi(x),
\ee
where $\lL$ is a constant vector from the $(-q)$-charged endpoint to the $(+q)$-charged endpoint of the dipole, and $x$ is taken at the center of the dipole. Such theories arise naturally in string theory and in the context of noncommutative geometry, as explained in \cite{Bergman:2000cw,Bergman:2001rw}.

The goal of this paper is to construct a 6d counterpart of the dipole-theories where the role of the gauge field $A_\mu dx^\mu$ is played by a gauge $2$-form $\Bz$ with anti-self-dual field strength $\Hz = -{}^*\Hz$ (and at $0^{th}$ order $\Hz=d\Bz$). The role of the dipole fields will then be played by ``curvepole'' fields.
Let us define a {\it curvepole} to be a non-pointlike particle that has the shape of a fixed oriented closed curve $\curveC\subset\R^6$ and that interacts with a given two-form field $\Bz=\tfrac{1}{2}\Bz_{\mu\nu}dx^\mu\wedge dx^\nu$ 
($\mu,\nu,\dots=1,\dots,6$ label the coordinates on Euclidean $\R^6$) via the action $i\int_{(x+\curveC)}\Bz$, where $(x+\curveC)$ denotes the result of translating $\curveC$ by the spacetime vector $x$. We would also like the role of the covariant derivative \eqref{eqn:defDdipole} to be played by \cite{Schaeffer}:
\be\label{eqn:defDcurvepoleB}
\partial_\mu\aSq(x)
+i\aSq(x)\int_\curveC\Bz_{\mu\nu}(x+y)dy^\nu.
\ee
However, \eqref{eqn:defDcurvepoleB} is incompatible with anti-self-duality and will require a small modification, to be explained in \ref{sec:construct}.

We also wish to preserve some amount of supersymmetry, and in fact, we will require $(1,0)$-SUSY. The two-form $\Bz$ will be part of a tensor multiplet (containing also a real scalar, and two Weyl spinors), and the curvepole field $\aSq$ will be part of a hyper-multiplet (containing two complex scalars and a complex Weyl spinor). The interaction also modifies the anti-self-duality relation $d\Bz=-{}^*d\Bz$, as we shall see.

Before proceeding to the details, let us discuss another reason to be interested in the curvepole deformation of the $(2,0)$-theory. As was suggested in \cite{Bergman:2000cw,Dasgupta:2000ry,Alishahiha:2003ru}, such theories arise naturally in M-theory.
Following \cite{Dasgupta:2000ry}, we probe with an M$2$-brane a flat M-theory background of the form $\R^{1,2}\times T^2\times\TransverseSpace{\Omega}$, where $\TransverseSpace{\Omega}$ is the $6$-dimensional flat space formed by fibering $\R^5$ [parameterized by $\vy=(x^6,\dots,x^{10})$] over $S^1$ (of radius $\rR$ and parameterized by $x^3$) with a twist $\Omega\in\text{Spin}(5)$. (I.e., we identify $x^3\simeq x^3+2\pi\rR$ together with $\vy\simeq\Omega\vy$.) For simplicity, we also take $T^2$ to be a product of two circles of radius $\rR$ (parameterized by $x^4,x^5$). We then set $\Omega = \exp(i\rh)$ [for $\rh$ a fixed element of the Lie algebra $so(5)$], insert an M$2$-brane probe at $x^3=\cdots=x^{10}=0$, and take the limit $\rh\rightarrow 0$, (Planck length) $\lP\rightarrow 0$, $\rR/\lP\rightarrow 0$, with $\rh\lP^4/\rR^2$ kept finite. 
Note that at the location of the M$2$-brane, directions $3,4,5$ form a small $T^3$ of volume $(2\pi\rR)^3\rightarrow 0$.
If also $\rh=0$, we can apply U-duality to turn the M$2$-brane into an M$5$-brane wrapped on a large $T^3$, which in turn is described at low-energy by a free tensor multiplet. It was argued in \cite{Dasgupta:2000ry,Alishahiha:2003ru} that for nonzero $\rh$ the dynamics has features of a curvepole theory. We will now proceed with a purely field theoretic construction, which we conjecture is applicable to the case that $\rh$ preserves half the supersymmetry, i.e., annihilates a spinor of $\text{Spin(5)}$.


\section{\label{sec:notat}Notation}

We work in Euclidean signature on $\R^6$ and denote coordinates by $x^\mu$ (with Greek indices taking values $\alpha,\beta,\dots = 1,\dots,6$).
Components of a $2$-form $B$ are denoted by $B_{\mu\nu}=-B_{\nu\mu}$ with $B=\tfrac{1}{2}B_{\mu\nu}dx^\mu\wedge dx^\nu$. The $3$-form field strength $H=dB$ has components $H_{\mu\nu\sigma} = 3\partial_{[\mu}B_{\nu\sigma]}$. Its self-dual $H^{(+)}_{\mu\nu\sigma}$ and anti-self-dual $H^{(-)}_{\mu\nu\sigma}$ parts are given by
$
H^{(\pm)}_{\mu\nu\sigma} = 
\tfrac{1}{2}
(H_{\mu\nu\sigma}
\pm\tfrac{i}{3!}\epsilon_{\mu\nu\sigma\alpha\beta\gamma}H^{\alpha\beta\gamma}),
$
where $\epsilon_{\alpha\beta\gamma\delta\mu\nu}$ is the Levi-Civita symbol.


\subsection{\label{subsec:spinor}Spinor conventions}

The 6d spin group is $\text{Spin(6)}\cong SU(4)$, and we use lowercase Roman indices ($\sa,\sb,\sc,\dots = 1,\dots,4$) to denote components of a chiral spinor, with lower indices for the fundamental representation $\rep{4}$ of $SU(4)$ (e.g., $\aPq_\sa$), and upper indices for the dual representation $\brep{4}$ (e.g., $\bPq^\sa$).
A vector in the representation $\rep{6}$ with components $V_\mu$ will be represented by an antisymmetric $V_{\sa\sb}=-V_{\sb\sa}$, normalized so that the scalar product of two vectors $V$ and $U$ is $V_\mu U^\mu = -\tfrac{1}{4}\epsilon^{\sa\sb\sc\sd}V_{\sa\sb}U_{\sc\sd}$, where $\epsilon^{\sa\sb\sc\sd}$ is totally antisymmetric with $\epsilon^{1234}=1$. We also define $V^{\sa\sb}\defineas\tfrac{1}{2}\epsilon^{\sa\sb\sc\sd}V_{\sc\sd}$ so that $V_\mu U^\mu = -\tfrac{1}{2}V_{\sa\sb}U^{\sa\sb}$.

Components of a $2$-form $B_{\mu\nu}$ are written in spinor notation as traceless $B_\sa^\sb$ (with $B_\sa^\sa=0$), and normalized so that $B_{\mu\nu}= U_{[\mu}V_{\nu]}$ translates to $B_\sa^\sb = 
\tfrac{1}{4}(V_{\sa\sc}U^{\sb\sc}-U_{\sa\sc}V^{\sb\sc}).$
We also note that $B_{\mu\nu}B^{\mu\nu}=-2B_\sa^\sb B_\sb^\sa$, and if $W^\mu$ and $T^\mu$ are vectors, then $T_\mu = B_{\mu\nu}W^\mu$ translates to $T^{\sa\sb} =2W^{\sc[\sa}B_\sc^{\sb]}$.

An anti-self-dual $3$-form with tensor components $H^{(-)}_{\mu\nu\sigma}$ is expressed in spinor notation as a symmetric $\Hz_{\sa\sb}=\Hz_{\sb\sa}$, while a self-dual $3$-form is expressed as $\Hz^{\sa\sb}=\Hz^{\sb\sa}$.
They are normalized so that $\Hz_{\sa\sb}\Hz^{\sa\sb}=\tfrac{1}{2}H_{\mu\nu\sigma}H^{\mu\nu\sigma}$. Moreover, $B_{\mu\nu}=H^{(-)}_{\mu\nu\sigma}V^\sigma$ translates to $B_\sa^\sb=-\tfrac{1}{2}H_{\sa\sc}V^{\sb\sc}$, and $B_{\mu\nu}=H^{(+)}_{\mu\nu\sigma}V^\sigma$ translates to $B_\sa^\sb=-\tfrac{1}{2}H^{\sb\sc}V_{\sa\sc}$. The relation $H_{\mu\nu\sigma} = 3\partial_{[\mu}B_{\nu\sigma]}$ translates to $\Hz_{\sa\sb}=2\partial_{\sc(\sa}B_{\sb)}^\sc$ and $\Hz^{\sa\sb}=2\partial^{\sc(\sa}B_\sc^{\sb)}$.


\subsection{\label{subsec:cpint}Curvepole integrals}

The action presented below depends on the choice of a fixed (two-dimensional) open surface $\SCP\subset\R^6$ with boundary $\curveC=\partial\SCP$. For a (either fundamental or composite) field $\Phi(x)$, we define the {\it curvepole integral}:
\be\label{eqn:defcpint}
\intSCP{\Phi(x)}^{\mu\nu}\defineas
\int_\SCP\Phi(x+y)dy^\mu\wedge dy^\nu
\,,
\ee
 and the {\it anti-curvepole integral}:
\be\label{eqn:defcprint}
\rintSCP{\Phi(x)}^{\mu\nu}\defineas
\int_\SCP\Phi(x-y)dy^\mu\wedge dy^\nu\,.
\ee
We call $\SCP$ the {\it curvepole surface}, and say that it is {\it balanced} if for arbitrary $\Phi$ and $\Psi$,
\be\label{eqn:balancedcp}
0 = \int\left(\intSCP{\Phi(x)}_{\alpha\beta}\intSCP{\Psi(x)}_{\gamma\delta}-
\intSCP{\Phi(x)}_{\gamma\delta}\intSCP{\Psi(x)}_{\alpha\beta}\right)d^6x
\,.
\ee
It can also be expressed as $\rintSCP{\intSCP{\Phi}_{\gamma\delta}}_{\alpha\beta}=\rintSCP{\intSCP{\Phi}_{\alpha\beta}}_{\gamma\delta}$.
We will call $\SCP$ {\it $n$-planar} if it can be embedded in some $\R^n$ subspace of $\R^6$.
It is not hard to see that a $2$-planar curvepole is balanced [in which case the integrand of \eqref{eqn:balancedcp} vanishes], but the reverse is not necessarily true. In fact, any parity 
invariant $\SCP$ is balanced.


\section{\label{sec:construct}The curvepole deformation}

We will start with the action $\AcI_0$ of the free 6d tensor theory, and add interaction terms $\AcI_1 +\AcI_2 + \cdots$, where $\AcI_n$ is a polynomial of order $(n+2)$ in the fundamental fields, and each of its terms contains a total of $n$ curvepole integrals. At each order we will modify the supersymmetry variation $\delta = \delta_0 + \delta_1 + \delta_2 + \cdots$, where $\delta_0$ is the free SUSY variation that is linear in the fields, $\delta_1$ is quadratic in the fields with terms that have one curvepole integral, etc. 
The calculations have been performed up to quartic order, so the results presented below satisfy
\be\label{eqn:deltaAll}
0 = \delta_0\AcI_0 = \delta_0\AcI_1 +\delta_1\AcI_0 =\delta_2\AcI_0+\delta_1\AcI_1+\delta_2\AcI_0\,.
\ee
The curvepole deformation preserves a $U(1)\times SU(2)$ subgroup of the $Sp(2)$ R-symmetry of the free tensor multiplet. $SU(2)$ doublets will carry indices $\ri,\rj,\rk,\dots = 1,2$. They will be lowered with the antisymmetric $\epsilon_{\ri\rj}$. The surviving anticommuting $(1,0)$ SUSY parameters, which are $U(1)$ neutral $SU(2)$ doublets and chiral spinors, will be denoted by $\pSUSY^{\sa\ri}$. The $U(1)$ symmetry will be further discussed in subsection \ref{subsec:order1}.


\subsection{\label{subsec:order0}Quadratic terms}

The fields of the $(2,0)$ tensor multiplet are: a $2$-form gauge field $\Bz_\sa^\sb$, an $SU(2)$ singlet scalar $\Sz$, an $SU(2)$ doublet anticommuting spinor $\Pz_\sa^\ri$,
an $SU(2)$ doublet scalar $\aSq^\ri$ and its complex conjugate $\bSq_\ri$, and an $SU(2)$ singlet anticommuting spinor $\aPq_\sa$ and its complex conjugate $\bPq_\sa$.
The first three fields form a $(1,0)$ tensor multiplet, and the last four form a hypermultiplet.
The SUSY transformations are
$$
\begin{array}{lcllcl}
\delta_0\aSq^\ri &=& \pSUSY^{\sa\ri}\aPq_\sa\,,
& 
\delta_0\aPq_\sa &=&
-2i\pSUSY^\sb_\ri\partial_{\sa\sb}\aSq^\ri\,,
\\
\delta_0\bSq_\ri &=& \pSUSY^\sa_\ri\bPq_\sa\,,
& 
\delta_0\bPq_\sa &=&
-2i\pSUSY^\sb_\ri\partial_{\sa\sb}\bSq^\ri\,,
\\
\delta \Bz_\sa^\sb &=&
\pSUSY^\sb_\ri\Pz_\sa^\ri
-\tfrac{1}{4}\delta_\sa^\sb\pSUSY^\sc_\ri\Pz_\sc^\ri\,,
& 
\delta_0\Sz &=& \pSUSY^\sa_\ri\Pz_\sa^\ri
\,,
\\
\delta_0\Pz^\ri_\sa &=&
i\pSUSY^{\sb\ri}\partial_{\sa\sb}\Sz
+i\pSUSY^{\sb\ri}\Hz_{\sa\sb}\,.
& & &
\\
\end{array}
$$
The action is
\bear
\AcI_0 &=&
\int d^6x\bigl(
\tfrac{1}{12\pi}\Hz_{\mu\nu\sigma}\Hz^{\mu\nu\sigma}
+\tfrac{1}{4\pi}\partial_\mu\Sz\partial^\mu\Sz
+\partial_\mu\bSq_\ri\partial^\mu\aSq^\ri
\nn\\ &&
-\tfrac{i}{2\pi}\Pz_{\ri\sa}\partial^{\sa\sb}\Pz^\ri_\sb
+\bPq_\sa\partial^{\sa\sb}\aPq_\sb
\bigr)\,.\label{eqn:I0}
\eear
Note that the coupling constant of the $3$-form field strength has the self-dual value $\sqrt{2\pi}$.
The self-dual part of the $3$-form field strength $\Hz^{(+)}_{\mu\nu\sigma}$ decouples, and we will make sure below that interaction terms are always expressed in terms of the anti-self-dual $\Hz^{(-)}_{\mu\nu\sigma}$ only.


\subsection{Cubic terms \label{subsec:order1}}

At $0^{th}$ order, the $U(1)\subset Sp(2)$ current that the curvepole deformation preserves is given by
\be\label{eqn:JTnew}
\JTnew_{\sa\sb}\defineas
i\bSq_\ri\partial_{\sa\sb}\aSq^\ri
-i\partial_{\sa\sb}\bSq_\ri\aSq^\ri
+\bPq_\sa\aPq_\sb
-\bPq_\sb\aPq_\sa\,.
\ee
The fields that will become curvepoles are those that are charged under this $U(1)$.
However, the covariant derivative \eqref{eqn:defDcurvepoleB} is not satisfactory, because interaction terms in our action are allowed to depend on $\Bz$ only through $\Hz^{(-)}$.
To overcome this, we recall that for dipole-theories there is a way to rewrite \eqref{eqn:defDdipole} so that only the field strength $F_{\mu\nu}=2\partial_{[\mu}A_{\nu]}$ appears \cite{Bergman:2000cw}. One performs a field redefinition $\newPhi(x)\defineas\exp\left(i q\int_{(x+\dPath)}A_\mu dx^\mu\right)\Phi(x)$, where $\dPath$ is some fixed path from $-\tfrac{1}{2}\lL$ to $\tfrac{1}{2}\lL$, and $x+\dPath$ is the translation of the path by $x$.
Then
$
e^{i q\int_{(x+\dPath)}A_\mu dx^\mu}
D_\mu^{\lbrack\lL\rbrack}\phi(x)
$
is
$$
\partial_\mu\newPhi(x)
-i q\newPhi(x)\int_\dPath F_{\mu\nu}(x+y)dy^\nu\,.
$$
For curvepole interactions, guided by the above example, we replace $q\int_\dPath F_{\mu\nu}(x+y)dy^\nu$ with $\pm\intSCP{\Hz^{(-)}_{\mu\nu\sigma}}^{\nu\sigma}$.
We thus add to $\AcI_0$ the interaction term $-\tfrac{1}{2}\JTnew^\mu\intSCP{\Hz^{(-)}_{\mu\nu\sigma}}^{\nu\sigma}$ and look for a supersymmetric completion. The results are described below.

We define a vector field 
\be\label{eqn:mVdef}
\mV_\mu\defineas
\tfrac{1}{2}\intSCP{\Hz^{(-)}_{\mu\nu\sigma}}^{\nu\sigma}
-\tfrac{1}{2}\intSCP{\partial^\nu\Sz}_{\mu\nu},
\ee
which will play the role of an effective gauge field. We also define an effective gluino field $\mR^{\ri\sa}$ and a composite field $\cSX^\ri_\sa$ by
\be\label{eqn:mRcSXdef}
\mR^{\ri\sa}\defineas
\intSCP{\partial^{\sa\sb}\Pz^\ri_\sc}^\sc_\sb
-\intSCP{\partial^{\sb\sc}\Pz^\ri_\sb}^\sa_\sc, 
\qquad
\cSX^\ri_\sa\defineas
\aSq^\ri\bPq_\sa+\bSq^\ri\aPq_\sa
.
\ee
Then, the cubic interactions in the action are given by
\bear
\AcI_1 &=&
\int d^6x\left(
-\JTnew^\mu\mV_\mu
+i\cSX_{\ri\sa}\mR^{\ri\sa}
\right)
\,.
\label{eqn:I1}
\eear
To be consistent with the equations of motion, the anti-self-duality condition is modified to
$$
H^{(+)}_{\mu\nu\sigma} = 
-\tfrac{3\pi i}{4}
\rintSCP{\JTnew_{[\mu}}_{\nu\sigma]}
+\tfrac{\pi}{8}\epsilon_{\mu\nu\sigma\alpha\beta\gamma}\rintSCP{\JTnew^\alpha}^{\beta\gamma}.
$$
To write the correction to the SUSY transformation at this order, we define
$\mL\defineas\pSUSY_\ri^\sa\intSCP{\Pz_\sc^\ri}^\sc_\sa$, and set $\delta_1\Sz = 0$, 
\be\label{eqn:del1q}
\left.
\begin{array}{lcllcl}
\delta_1\aSq^\ri &=& i\mL\aSq^\ri\,,
& 
\delta_1\aPq_\sa  &=&
2\pSUSY^\sb_\ri\mV_{\sa\sb}\aSq^\ri +i\mL\aPq_\sa
\,,\\
\delta_1\bSq^\ri &=& -i\mL\bSq^\ri
\,,
&
\delta_1\bPq_\sa &=&
-2\pSUSY^\sb_\ri\mV_{\sa\sb}\bSq^\ri -i\mL\bPq_\sa
\,,
\\
\end{array}
\right\}
\ee
\bear
\lefteqn{
\delta_1\Pz^\ri_\sa =
\tfrac{\pi i}{2}\pSUSY^{\sc\ri}\rintSCP{\bPq_\sb\aPq_\sa-\bPq_\sa\aPq_\sb}^\sb_\sc
+\tfrac{\pi i}{2}\pSUSY^{\sc\ri}\rintSCP{\bPq_\sb\aPq_\sc-\bPq_\sc\aPq_\sb}^\sb_\sa
}\nn\\ &&
-\tfrac{\pi}{2}\pSUSY^\sc_\rj\rintSCP{3(\bSq^\ri\partial_{\sb\sc}\aSq^\rj
+\aSq^\ri\partial_{\sb\sc}\bSq^\rj)
+\bSq^\rj\partial_{\sb\sc}\aSq^\ri
+\aSq^\rj\partial_{\sb\sc}\bSq^\ri}^\sb_\sa
\nn\\ && 
+\tfrac{\pi}{2}\pSUSY^\sc_\rj\rintSCP{\aSq^\ri\partial_{\sb\sa}\bSq^\rj
+\bSq^\ri\partial_{\sb\sa}\aSq^\rj+3(\aSq^\rj\partial_{\sb\sa}\bSq^\ri
+\bSq^\rj\partial_{\sb\sa}\aSq^\ri)}^\sb_\sc
\,,
\nn
\eear
\bear
\delta_1B^\sa_\sb &=&
\pi i\pSUSY^\sa_\ri\rintSCP{\cSX^\ri_\sc}^\sc_\sb
-\tfrac{\pi i}{2}\pSUSY^\sc_\ri
\left(\rintSCP{\cSX^\ri_\sc}^\sa_\sb
-2\rintSCP{\cSX^\ri_\sb}^\sa_\sc
+\delta_{\sb}^{\sa}\rintSCP{\cSX^\ri_\sd}^\sd_\sc\right)
.\nn
\eear


\subsection{\label{subsec:order2}Quartic terms}

We define an $SU(2)$ triplet ``meson'' composite field
$$
\QT^\rj_\ri\defineas\bSq_\ri\aSq^\rj-\tfrac{1}{2}\delta_\ri^\rj\bSq_\rk\aSq^\rk\,,\qquad
(\QT^\ri_\ri=0).
$$
We then find that 
\bear
\lefteqn{
\AcI_2 =
\int d^6x\bigl(
\mV_\mu\mV^\mu\bSq_\ri\aSq^\ri
}\nn\\ &&
-\tfrac{3\pi}{16}\intSCP{\JTnew^{[\mu}}^{\sigma\nu]}\intSCP{\JTnew_{[\mu}}_{\sigma\nu]}
+\tfrac{\pi}{2}\intSCP{\partial^\mu\QT^\rj_\ri}_{\mu\sigma}
\intSCP{\partial_\nu\QT^\ri_\rj}^{\nu\sigma}
\bigr)
\label{eqn:I2}
\eear
satisfies \eqref{eqn:deltaAll} provided the {\it balanced curvepole} condition 
\eqref{eqn:balancedcp} holds.
(We will not present $\delta_2$ here, but we claim that $\delta_0\AcI_2 + \delta_1\AcI_1$ vanishes up to $0^{th}$ order equations of motion, which is equivalent to the existence of a $\delta_2$ such that $\delta_0\AcI_2+\delta_1\AcI_1+\delta_2\AcI_0=0$.)
The balanced curvepole condition is needed to cancel terms of the form $\intSCP{\bSq\partial\aPq}\intSCP{\bPq\aPq}$.

The first term $\mV_\mu\mV^\mu\bSq_\ri\aSq^\ri$ in \eqref{eqn:I2} combines with $-\JTnew^\mu\mV_\mu$ of \eqref{eqn:I1} and the kinetic term of \eqref{eqn:I0} to give the action of a scalar that is minimally coupled to the effective gauge field $\mV_\mu$. The second term of \eqref{eqn:I2} is part of an effective magnetic coupling. It can also be derived by formally dualizing an electrically coupled $\Hz_{\mu\nu\sigma}$ using the standard technique of treating $\Hz$ as the independent field and introducing a Lagrange multiplier for the Bianchi identity $d\Hz=0$, then integrating over $\Hz$. (See \cite{Witten:1996hc} for a thorough explanation on how to couple a theory of anti-self-dual fields to a $3$-form source,
and see \cite{Ganor:2000my,Gopakumar:2000na,Ganor:2002ju} for related calculations in the context of other nonlocal gauge theories.)

The terms in $\AcI_2$ were determined by requiring supersymmetry up to (and including) quartic terms. But SUSY alone is not sufficient to uniquely determine $\AcI_2$, because it turns out that there is a term
$
\AcI_2' =
\int d^6x\bigl(
\tfrac{1}{2}\intSCP{\JTnew_\sigma}_{\mu\nu}\intSCP{\JTnew^\sigma}^{\mu\nu}
-\intSCP{\partial_\sigma\QT^\ri_\rj}_{\mu\nu}\intSCP{\partial^\sigma\QT^\rj_\ri}^{\mu\nu}
+i\intSCP{\cSX^\ri_\sa}_{\mu\nu}\partial^{\sa\sb}\intSCP{\cSX_{\ri\sb}}^{\mu\nu}
\bigr)
$
that satisfies $\delta_0\AcI_2'+\delta_2'\AcI_0$ for an appropriate $\delta_2'$.
Thus, $\AcI_2\rightarrow\AcI_2+c\AcI_2'$, together with $\delta_2\rightarrow\delta_2+c\delta_2'$, will also be a solution to \eqref{eqn:deltaAll} for any arbitrary constant $c$. Nevertheless, \eqref{eqn:I2} is uniquely determined by imposing the ``hollowness'' requirement to be introduced next.


\section{\label{sec:hollow}Hollow curvepoles}

We would like the curvepole theory defined by $\AcI=\AcI_0+\AcI_1+\AcI_2+\cdots$ to depend only on the curve $\curveC=\partial\SCP$, and not on the way we fill it with the curvepole surface $\SCP$. We will refer to terms that only depend on $\partial\SCP$ as {\it hollow.}
As it stands, however, \eqref{eqn:I1} and \eqref{eqn:I2} appear to depend on the bulk of $\SCP$. Upon closer inspection, we see that the last term of \eqref{eqn:I2} and the last term of \eqref{eqn:I1} in fact do depend only on $\curveC$. This is because
$
\intSCP{\partial_\nu \Phi}^{\nu\sigma}
=\int_\curveC\Phi(x+y)dy^\sigma,
$
and therefore $\intSCP{\partial_\nu\QT^\ri_\rj}^{\nu\sigma}$ only depends on $\curveC$. 
Moreover, $\mR^{\ri\sa}$, which was defined in \eqref{eqn:mRcSXdef}, can be expressed as the contraction $\sc=\sd$ of the expression
$\intSCP{\partial^{\sa\sb}\Pz^\ri_\sd}^\sc_\sb
-\intSCP{\partial^{\sc\sb}\Pz^\ri_\sd}^\sa_\sb$,
and for fixed $\sd$, this is antisymmetric in $\sa,\sb$, and hence a vector that is proportional to $\intSCP{\partial_\mu\Pz^\ri_\sd}^{\mu\nu}$, which depends only on $\curveC$. The terms in the action that are not explicitly hollow are
\bear
\lefteqn{
\int d^6x\Bigl(
-\tfrac{1}{2}J^\mu\intSCP{\Hz_{\mu\nu\sigma}^{(-)}}^{\nu\sigma}
}\nn\\ &&
\qquad\qquad
+\mV_\mu\mV^\mu\bSq_\ri\aSq^\ri
-\tfrac{3\pi}{16}\intSCP{\JTnew^{[\mu}}^{\sigma\nu]}\intSCP{\JTnew_{[\mu}}_{\sigma\nu]}
\Bigr).
\label{eqn:nonhollow}
\eear
The first term of \eqref{eqn:nonhollow} could have been hollow if $\Hz^{(-)}$ were a closed $3$-form. But this is only true at $0^{th}$ order. The equations of motion that follow from $\AcI_0+\AcI_1$ are
\bear
\partial_{[\alpha}\Hz^{(-)}_{\mu\nu\sigma]} &=&
-\tfrac{3\pi i}{32}\epsilon_{\alpha\mu\nu\sigma\gamma\delta}\rintSCP{\partial_\tau\JTnew^{[\gamma}}^{\delta\tau]}
+\tfrac{3\pi}{4}\rintSCP{\partial_{[\alpha}\JTnew_\mu}_{\nu\sigma]}.
\qquad
\label{eqn:dHz1}
\eear
However, it turns out that the total action is hollow up to a field redefinition, provided $\SCP$ is $3$-planar.
To see this, consider a deformation of $\SCP$ given by $y^\mu\rightarrow y^\mu+\yV^\mu$,
 where $\yV:\SCP\rightarrow\R^6$ is an infinitesimal vector that vanishes on $\curveC$.
Then, for any field $\Phi$, the variation of the curvepole integral is
\be\label{eqn:deltayVint}
\delta_\yV\intSCP{\Phi}^{\mu\nu}
=
\intSCP{\yV^\sigma\partial_\sigma\Phi}^{\mu\nu}
+\intSCP{\yV^\mu\partial_\sigma\Phi}^{\nu\sigma}
-\intSCP{\yV^\nu\partial_\sigma\Phi}^{\mu\sigma}
,
\ee
where $\intSCP{\yV^\sigma(\cdots)}$ is defined as the integral \eqref{eqn:defcpint} with $\yV^\sigma(y)$ inserted in the integrand, and similarly $\rintSCP{\yV^\sigma(\cdots)}$ is given by \eqref{eqn:defcprint} with $\yV^\sigma(-y)$.
Next, we augment \eqref{eqn:deltayVint} with the field redefinition
$
\delta\Bz_{\gamma\delta} = \tfrac{\pi i}{4}
\epsilon_{\alpha\mu\nu\sigma\gamma\delta}\rintSCP{\yV^\alpha\JTnew^\mu}^{\nu\sigma}
$
(note that $i$ will be absent in Minkowski signature),
as well as
$\delta\aSq^\ri = i\varepsilon\aSq^\ri$,
$\delta\bSq^\ri = -i\varepsilon\bSq^\ri$,
$\delta\aPq^\ri = i\varepsilon\aPq^\ri$,
$\delta\bPq^\ri = -i\varepsilon\bPq^\ri$,
where $\varepsilon\defineas\tfrac{1}{2}\intSCP{\yV^\alpha\Hz^{(-)}_{\alpha\mu\nu}}^{\mu\nu}$.
Then a straightforward calculation shows that
\be\label{eqn:dyV}
\delta_\yV\bigl(\AcI_0 + \AcI_1 + \AcI_2\bigr)
=
\tfrac{\pi i}{32}\int\epsilon_{\alpha\beta\gamma\mu\nu\sigma}\intSCP{\JTnew^\alpha}^{\beta\gamma}
\intSCP{\yV^\tau\partial_\tau\JTnew^\mu}^{\nu\sigma} d^6x,
\ee
The RHS of \eqref{eqn:dyV} vanishes for a $3$-planar $\SCP$, because in a suitable coordinate system there are only three possible values for the indices $\beta,\gamma,\nu,\sigma$ for which the curvepole integral does not vanish. Thus, if $\curveC$ is $2$-planar, we can deform $\SCP$ in the direction of an arbitrary $3$-plane without affecting the theory. This demonstrates that the bulk of $\SCP$ is ``incorporeal.''
Recall that a $2$-planar curvepole $\SCP$ is balanced. We now see that even if only $\curveC$ is $2$-planar, as long as $\SCP$ is $3$-planar the theory is supersymmetric (up to quartic order), although perhaps not manifestly so.


\section{\label{sec:disc}Discussion}

We have constructed, up to quartic order, a supersymmetric Lagrangian that describes the interactions of a curvepole field with a tensor gauge field $B_{\mu\nu}$ whose field strength is anti-self-dual.
Requiring, in addition to supersymmetry, that the interactions be only through the boundary $\curveC=\partial\SCP$, in the sense of section \ref{sec:hollow}, determines the quartic terms uniquely.

The question of quintic and higher terms is open. Terms of order $n$ (within $\AcI_{n-2}$) have a total of $(n-2)$ $\intSCP{\cdot}$'s and $\rintSCP{\cdot}$'s.
Basic dimensional analysis shows that no such terms are explicitly hollow for $n\ge 5$ (since each $\intSCP{\cdot}$ would require an additional $\partial$ in order to be expressible as an integral over $\curveC$). Thus, to preserve ``hollowness'', we would expect the only modification to be that $\JTnew^\mu$ in \eqref{eqn:I2} should be corrected from its $0^{th}$ order form \eqref{eqn:JTnew} to include the contribution of interactions. It is easy to derive it order by order as the Noether current of the $U(1)$ symmetry.

At order $n=6$ possible anomalies should also be considered.
The SUSY transformations \eqref{eqn:del1q} involve anomalous chiral rotations of fermions. However, since $\mV$ is not a true gauge field, this is not expected to be a problem.
In fact, without the chiral rotation the SUSY algebra closes only up to an anomalous $U(1)$ gauge transformation, and hence supersymmetry would be anomalous itself (see \cite{Riccioni:1998th}). The chiral rotation ensures that the anticommutator of two SUSY transformations is a translation, and hence is expected to be anomaly free.
The fermionic field redefinitions of section \ref{sec:hollow} also have an anomaly proportional to $\int\varepsilon(d\mV)^3$, and it is possible that the ``hollowness'' condition suffers from anomalies at $n\ge 6$. An anomaly would manifest itself as a nonzero $\delta_\yV$ variation of a $4$-point correlator of tensor fields, of the form $\langle \Hz\Hz\Hz\Hz\rangle$. This variation, however, also receives a $1$-loop contribution from 
the field redefinition $
\delta\Bz_{\gamma\delta} = \tfrac{\pi i}{4}
\epsilon_{\alpha\mu\nu\sigma\gamma\delta}\rintSCP{\yV^\alpha\JTnew^\mu}^{\nu\sigma}
$, introduced in \ref{sec:hollow}. The complete result, and in particular the question of whether a hollowness-anomaly exists or not, would require further analysis.

We expect the action constructed here to describe the low-energy degrees of freedom of the M-theory setup described at the end of section \ref{sec:intro} when the twist $\rh$ has a nonzero kernel in the spinor representation (and hence preserves SUSY). It would be interesting to determine $\SCP$ from the twist.
That setup has a formal dual with an M$5$-brane probing a background with strong $4$-form flux \cite{Bergman:2001rw,Alishahiha:2003ru}. The limit of small $\SCP$ might therefore be related to the order-by-order expansion of the interaction of an M$5$-brane with $4$-form flux (see also \cite{Aganagic:1997zq,Bergshoeff:2000jn,Gopakumar:2000ep,Lambert:2014fma,Lambert:2016xbs,Lambert:NewPaper}).
It would also be interesting to connect the M-theory construction of \cite{Bergman:2000cw,Alishahiha:2003ru} with the field theory construction of \cite{Ho:2008nn} (using BLG theory \cite{Bagger:2006sk,Gustavsson:2007vu}), and to generalize to other 6d theories \cite{Ganor:1996mu,Seiberg:1996vs,Blum:1997mm,DelZotto:2014hpa,Heckman:2015bfa}.

Curvepole theories can be viewed as higher dimensional analogs of dipole-theories, and the latter have interesting supergravity duals \cite{Bergman:2001rw} that have found applications in the study of rotating black holes \cite{Song:2011sr,ElShowk:2011cm}.
We would expect a curvepole deformation to exist for any number of M$5$-branes, and it would be interesting to identify deformations of $AdS_7\times S^4$ \cite{Maldacena:1997re} that depend on $\SCP$ as a parameter.

Curvepole theories can also serve as novel kinds of UV completion of 5d theories.
Consider curvepole theory on $\R^5\times S^1$, with $\SCP=I\times S^1$, where $I\subset\R^5$ is a straight segment. $\SCP$ is $2$-planar and hence balanced. We can take one end of the segment to $\infty$ and the other end at the origin. According to section \ref{sec:hollow}, it does not matter in which direction we take the segment.
The theory is therefore local on $\R^5$, and at low energy it becomes a $U(1)$ gauge field interacting with a charged hyper-multiplet. (See \cite{Tachikawa:2015mha} for a recent analysis of UV completions of supersymmetric 5d theories.)

\begin{acknowledgments}
I am grateful to Kevin Schaeffer for a discussion on \cite{Bergman:2000cw,Dasgupta:2000ry,Alishahiha:2003ru} in 2012, in which he explicitly wrote down the covariant derivative form \eqref{eqn:defDcurvepoleB}, and which subsequently prompted me to search for a supersymmetric extension. I would also like to thank Keshav Dasgupta for helpful comments, Kuo-Wei Huang for helpful correspondence, Neil Lambert for sharing with me the yet unpublished results of \cite{Lambert:NewPaper}, and the Aspen Center for Theoretical Physics and the organizers of the 2017 Aspen Winter Conference ``Superconformal Field Theories in $d\ge4$'', for allowing me to present preliminary results of this work.

\end{acknowledgments}
%

\end{document}